\documentclass[11pt,epsfig,psfig]{article}
\usepackage{amssymb,amsmath,amsfonts}
\usepackage{graphicx}
\usepackage{graphics}
\usepackage{eepic,epsfig}

\begin{document}

\title{Casimir densities for wedge-shaped boundaries}
\author{A. A. Saharian\thanks{%
E-mail: saharian@ictp.it} \\
Department of Physics, Yerevan State University \\
1 Alex Manoogian Street, 0025 Yerevan, Armenia }
\date{ }
\maketitle

\begin{abstract}
The vacuum expectation values of the field squared and the energy-momentum
tensor are investigated for a scalar field with Dirichlet boundary
conditions and for the electromagnetic field inside a wedge with a coaxial
cylindrical boundary. In the case of the electromagnetic field perfectly
conducting boundary conditions are assumed on the bounding surfaces. By
using the Abel-Plana-type formula for the series over the zeros of the
Bessel function, the vacuum expectation values are presented in the form of
the sum of two terms. The first one corresponds to the geometry without a
cylindrical boundary and the second one is induced by the presence of the
cylindrical shell. The additional vacuum forces acting on the wedge sides
due the presence of the cylindrical boundary are evaluated and it is shown
that these forces are attractive for both scalar and electromagnetic fields.
\end{abstract}

\bigskip

PACS numbers: 11.10.Kk, 03.70.+k

\bigskip

\section{Introduction}

The Casimir effect has important implications on all scales, from
cosmological to subnuclear, and has become in recent decades an increasingly
popular topic in quantum field theory. Since the original work by Casimir
\cite{Casi48} many theoretical and experimental works have been done on this
problem (see, e.g., \cite{Most97,Eliz94} and references therein). In
particular, a great deal of attention received the investigations of quantum
effects for cylindrical boundaries. In addition to traditional problems of
quantum electrodynamics under the presence of material boundaries, the
Casimir effect for cylindrical geometries can also be important to the flux
tube models of confinement \cite{Fish87} and for determining the structure
of the vacuum state in interacting field theories \cite{Ambj83}. The
calculation of the vacuum energy of electromagnetic field with boundary
conditions defined on a cylinder turned out to be technically a more
involved problem than the analogous one for a sphere. First the Casimir
energy of an infinite perfectly conducting cylindrical shell has been
calculated in Ref. \cite{Dera81} by introducing ultraviolet cutoff and later
the corresponding result was derived by zeta function technique \cite{Milt99}
(for a recent discussion of the Casimir energy and self-stresses in the more
general case of a dielectric-diamagnetic cylinder see \cite{Milt06} and
references therein). The local characteristics of the corresponding
electromagnetic vacuum such as energy density and vacuum stresses are
considered in \cite{Sah1cyl} for the interior and exterior regions of a
conducting cylindrical shell, and in \cite{Sah2cyl} for the region between
two coaxial shells (see also \cite{SahRev1}). The vacuum forces acting on
the boundaries in the geometry of two cylinders are also considered in Refs.
\cite{Mazz02}. The scalar Casimir densities for a single and two coaxial
cylindrical shells with Robin boundary conditions are investigated in Refs.
\cite{Rome01,Saha06}. Less symmetric configuration of two eccentric
perfectly conducting cylinders is considered in \cite{Mazz02}. Vacuum energy
for a perfectly conducting cylinder of elliptical section is evaluated in
Ref. \cite{Kits06} by the mode summation method, using the ellipticity as a
perturbation parameter. The Casimir forces acting on two parallel plates
inside a conducting cylindrical shell are investigated in Ref. \cite{Mara07}%
. The Casimir effect in more complicated geometries with cylindrical
boundaries is considered in \cite{Cascyl}.

Aside from their own theoretical and experimental interest, the problems
with this type of boundaries are useful for testing the validity of various
approximations used to deal with more complicated geometries. From this
point of view the wedge with a coaxial cylindrical boundary is an
interesting system, since the geometry is nontrivial and it includes two
dynamical parameters, radius of the cylindrical shell and opening angle of
the wedge. This geometry is also interesting from the point of view of
general analysis for surface divergences in the expectation values of local
physical observables for boundaries with discontinuities. The nonsmoothness
of the boundary generates additional contributions to the heat kernel
coefficients (see, for instance, the discussion in \cite{Apps98} and
references therein). In the present paper we review the results of the
investigations for the vacuum expectation values of the field squared and
the energy-momentum tensor for the scalar and electromagnetic fields in the
geometry of a wedge with a coaxial cylindrical boundary. In addition to
describing the physical structure of the quantum field at a given point, the
energy-momentum tensor acts as the source of gravity in the Einstein
equations. It therefore plays an important role in modelling a
self-consistent dynamics involving the gravitational field. Some most
relevant investigations to the present paper are contained in Refs. \cite%
{Most97,jphy,Deutsch,brevikI,Brev98,brevikII,Nest02}, where the geometry of
a wedge without a cylindrical boundary is considered for a conformally
coupled scalar and electromagnetic fields in a four dimensional spacetime.
The Casimir effect in open geometries with edges is investigated in \cite%
{Gies06}. The total Casimir energy of a semi-circular infinite cylindrical
shell with perfectly conducting walls is considered in \cite{Nest01} by
using the zeta function technique. The Casimir energy for the wedge-arc
geometry in two dimensions is discussed in \cite{Kolo07}. For a scalar field
with an arbitrary curvature coupling parameter the Wightman function, the
vacuum expectation values of the field squared and the energy-momentum
tensor in the geometry of a wedge with an arbitrary opening angle and with a
cylindrical boundary are evaluated in \cite{Reza02,Saha05}. Note that,
unlike the case of conformally coupled fields, for a general coupling the
vacuum energy-momentum tensor is angle-dependent and diverges on the wedge
sides. The corresponding problem for the electromagnetic field, assuming
that all boundaries are perfectly conducting, is investigated in \cite%
{Saha07}. The scalar Casimir densities in the geometry of a wedge with two
cylindrical boundaries are discussed in \cite{Saha08}. The closely related
problem of the vacuum densities induced by a cylindrical boundary in the
geometry of a cosmic string is investigated in Refs. \cite{Beze06} for
scalar, electromagnetic and fermionic fields.

We have organized the paper as follows. The next section is devoted to the
evaluation of the Wightman function for a scalar field with a general
curvature coupling inside a wedge with a cylindrical boundary. By using the
formula for the Wightman function, in section \ref{sec:Wedgecylabs} we
evaluate the vacuum expectation values of the field squared and the
energy-momentum tensor inside a wedge without a cylindrical boundary. The
vacuum densities for a wedge with the cylindrical shell are considered in
section \ref{sec:Wedgecyl}. Formulae for the shell contributions are derived
and the corresponding surface divergences are investigated. The vacuum
expectation values of the electric and magnetic field squared inside a wedge
with a cylindrical boundary are investigated in section \ref{sec:Inter},
assuming that all boundaries are perfectly conducting. The corresponding
expectation values for the electromagnetic energy-momentum tensor are
considered in section \ref{sec:EMTint}. The results are summarized in
section \ref{sec:Conc}.

\section{Wightman function for a scalar field}

\label{sec:Wedge}

\begin{figure}[tbph]
\begin{center}
\epsfig{figure=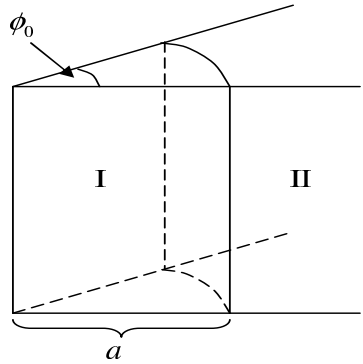, width=5cm, height=5cm}
\end{center}
\caption{Geometry of a wedge with the opening angle $\protect\phi _0$ and
cylindrical boundary of radius $a$.}
\label{fig1}
\end{figure}

Consider a free scalar field $\varphi (x)$ inside a wedge with the opening
angle $\phi _{0}$ and with a cylindrical boundary of radius $a$ (see figure %
\ref{fig1}). We will use cylindrical coordinates $(x^{1},x^{2},\ldots
,x^{D})=(r,\phi ,z_{1},\ldots ,z_{N})$, $N=D-2$, where $D$ is the number of
spatial dimensions. The field equation has the form
\begin{equation}
\left( \nabla ^{i}\nabla _{i}+m^{2}+\xi R\right) \varphi (x)=0,
\label{fieldeq}
\end{equation}%
where $R$ is the scalar curvature for the background spacetime and $\xi $ is
the curvature coupling parameter. The special cases $\xi =0$ and $\xi =\xi
_{c}=(D-1)/4D$ correspond to minimally and conformally coupled scalars
respectively.

In this section we evaluate the positive frequency Wightman function $%
\langle 0|\varphi (x)\varphi (x^{\prime })|0\rangle $ assuming that the
field obeys Dirichlet boundary condition on the bounding surfaces:
\begin{equation}
\varphi |_{\phi =0}=\varphi |_{\phi =\phi _{0}}=\varphi |_{r=a}=0.
\label{Dirbc}
\end{equation}%
The vacuum expectation value (VEV) of the energy-momentum tensor is
expressed in terms of the Wightman function as%
\begin{equation}
\langle 0|T_{ik}(x)|0\rangle =\lim_{x^{\prime }\rightarrow x}\nabla
_{i}\nabla _{k}^{\prime }\langle 0|\varphi (x)\varphi (x^{\prime })|0\rangle
+\left[ \left( \xi -\frac{1}{4}\right) g_{ik}\nabla ^{l}\nabla _{l}-\xi
\nabla _{i}\nabla _{k}\right] \langle 0|\varphi ^{2}(x)|0\rangle .
\label{vevEMTWf}
\end{equation}%
In addition, the response of a particle detector in an arbitrary state of
motion is determined by this function. In (\ref{vevEMTWf}) we have assumed
that the background spacetime is flat and the term with the Ricci tensor is
omitted. The Wightman function is presented as the mode sum%
\begin{equation}
\langle 0|\varphi (x)\varphi (x^{\prime })|0\rangle =\sum_{\mathbf{\alpha }%
}\varphi _{\mathbf{\alpha }}(x)\varphi _{\mathbf{\alpha }}^{\ast }(x^{\prime
}),  \label{vevWf}
\end{equation}%
where $\{\varphi _{\mathbf{\alpha }}(x),\varphi _{\mathbf{\alpha }}^{\ast
}(x)\}$ is a complete orthonormal set of solutions to the field equation,
satisfying the boundary conditions, $\alpha $ is a set of the corresponding
quantum numbers.

\subsection{Interior region}

In the region $0\leqslant r\leqslant a$ (region I in figure \ref{fig1}), the
eigenfunctions satisfying the boundary conditions (\ref{Dirbc}) on the wedge
sides $\phi =0,\phi _{0}$ have the form
\begin{eqnarray}
\varphi _{\alpha }(x) &=&\beta _{\alpha }J_{qn}(\gamma r)\sin (qn\phi )\exp
\left( i\mathbf{kr}_{\parallel }-i\omega t\right) ,  \label{eigfunccirc} \\
\omega &=&\sqrt{\gamma ^{2}+k_{m}^{2}},\;k_{m}^{2}=|\mathbf{k}%
|^{2}+m^{2},\quad q=\pi /\phi _{0},  \label{qu}
\end{eqnarray}%
where $\alpha =(n,\gamma ,\mathbf{k})$, $-\infty <k_{j}<\infty $, $%
n=1,2,\cdots $, $\mathbf{k}=(k_{1},\ldots ,k_{N})$, $\mathbf{r}_{\parallel
}=(z_{1},\ldots ,z_{N})$, and $J_{l}(z)$ is the Bessel function. The
normalization coefficient $\beta _{\alpha }$ is determined from the standard
Klein-Gordon scalar product with the integration over the region inside the
wedge and is equal to
\begin{equation}
\beta _{\alpha }^{2}=\frac{2}{(2\pi )^{N}\omega \phi _{0}a^{2}J_{qn}^{\prime
2}(\gamma a)}.  \label{betalf}
\end{equation}%
The eigenvalues for the quantum number $\gamma $ are quantized by the
boundary condition (\ref{Dirbc}) on the cylindrical surface $r=a$. From this
condition it follows that
\begin{equation}
\gamma =\lambda _{n,j}/a,\quad j=1,2,\cdots ,  \label{ganval}
\end{equation}%
where $\lambda _{n,j}$ are the positive zeros of the Bessel function, $%
J_{qn}(\lambda _{n,j})=0$, arranged in ascending order, $\lambda
_{n,j}<\lambda _{n,j+1}$.

Substituting the eigenfunctions (\ref{eigfunccirc}) into mode sum formula (%
\ref{vevWf}) with the set of quantum numbers $\alpha =(n,j,\mathbf{k})$, for
the positive frequency Wightman function one finds
\begin{equation}
\langle 0|\varphi (x)\varphi (x^{\prime })|0\rangle =\int d^{N}\mathbf{k}%
\,e^{i\mathbf{k}\Delta \mathbf{r}_{\parallel }}\sum_{n=1}^{\infty }\sin
(qn\phi )\sin (qn\phi ^{\prime })\sum_{j=1}^{\infty }\beta _{\alpha
}^{2}J_{qn}(\gamma r)J_{qn}(\gamma r^{\prime })e^{-i\omega \Delta t},
\label{Wf1}
\end{equation}%
where $\gamma =\lambda _{n,j}/a$, and $\Delta \mathbf{r}_{\parallel }=%
\mathbf{r}_{\parallel }-\mathbf{r}_{\parallel }^{\prime }$, $\Delta
t=t-t^{\prime }$. In order to obtain an alternative form for the Wightman
function we apply to the sum over $j$ a variant of the generalized
Abel-Plana summation formula \cite{Sahrev}
\begin{eqnarray}
\sum_{j=1}^{\infty }\frac{2f(\lambda _{n,j})}{\lambda _{n,j}J_{qn}^{\prime
2}(\lambda _{n,j})} &=&\int_{0}^{\infty }f(z)dz+\frac{\pi }{4}\mathrm{Res}%
_{z=0}\left[ f(z)\frac{Y_{qn}(z)}{J_{qn}(z)}\right] -  \notag \\
&&+\frac{i}{\pi }\int_{0}^{\infty }dz\,\frac{K_{qn}(z)}{I_{qn}(z)}[e^{-qn\pi
i}f(ze^{\pi i/2})-e^{qn\pi i}f(ze^{-\pi i/2})],  \label{sumform1AP}
\end{eqnarray}%
where $Y_{l}(z)$ is the Neumann function, and $I_{l}(z)$, $K_{l}(z)$ are the
modified Bessel functions. The corresponding conditions for the formula (\ref%
{sumform1AP}) to be valid are satisfied if $r+r^{\prime }+|t-t^{\prime }|<2a$%
. In particular, this is the case in the coincidence limit $t=t^{\prime }$
for the region under consideration, $r,r^{\prime }<a$. Formula (\ref%
{sumform1AP}) allows to present the Wightman function in the form%
\begin{equation}
\langle 0|\varphi (x)\varphi (x^{\prime })|0\rangle =\langle 0_{w}|\varphi
(x)\varphi (x^{\prime })|0_{w}\rangle +\langle \varphi (x)\varphi (x^{\prime
})\rangle _{\mathrm{cyl}},  \label{Wf2}
\end{equation}%
where%
\begin{eqnarray}
\langle 0_{w}|\varphi (x)\varphi (x^{\prime })|0_{w}\rangle &=&\frac{1}{\phi
_{0}}\int \frac{d^{N}\mathbf{k}}{(2\pi )^{N}}e^{i\mathbf{k}\Delta \mathbf{r}%
_{\parallel }}\int_{0}^{\infty }dz\frac{ze^{-i\Delta t\sqrt{z^{2}+k_{m}^{2}}}%
}{\sqrt{z^{2}+k_{m}^{2}}}  \notag \\
&&\times \sum_{n=1}^{\infty }\sin (qn\phi )\sin (qn\phi ^{\prime
})J_{qn}(zr)J_{qn}(zr^{\prime }),  \label{Wf00}
\end{eqnarray}%
and%
\begin{eqnarray}
\langle \varphi (x)\varphi (x^{\prime })\rangle _{\mathrm{cyl}} &=&-\frac{2}{%
\pi \phi _{0}}\int \frac{d^{N}\mathbf{k}}{(2\pi )^{N}}e^{i\mathbf{k}\Delta
\mathbf{r}_{\parallel }}\int_{k_{m}}^{\infty }dz\frac{z\cosh (\Delta t\sqrt{%
z^{2}-k_{m}^{2}})}{\sqrt{z^{2}-k_{m}^{2}}}  \notag \\
&&\times \sum_{n=1}^{\infty }\sin (qn\phi )\sin (qn\phi ^{\prime
})I_{qn}(zr)I_{qn}(zr^{\prime })\frac{K_{qn}(za)}{I_{qn}(za)}.  \label{Wfa0}
\end{eqnarray}%
In the limit $a\rightarrow \infty $ for fixed $r,r^{\prime }$, the term $%
\langle \varphi (x)\varphi (x^{\prime })\rangle _{\mathrm{cyl}}$ vanishes
whereas the part (\ref{Wf00}) does not depend on $a$. Hence, the term $%
\langle 0_{w}|\varphi (x)\varphi (x^{\prime })|0_{w}\rangle $ is the
Wightman function for the wedge without a cylindrical boundary with the
corresponding vacuum state $|0_{w}\rangle $. Consequently, the term $\langle
\varphi (x)\varphi (x^{\prime })\rangle _{\mathrm{cyl}}$ is induced by the
presence of the cylindrical boundary. For points away the cylindrical
surface this part is finite in the coincidence limit and the renormalization
is needed only for the part coming from the term (\ref{Wf00}).

\subsection{Exterior region}

In the region outside the cylindrical shell (region II in figure \ref{fig1}%
): $r>a$, $0\leqslant \phi \leqslant \phi _{0}$, the eigenfunctions
satisfying boundary conditions (\ref{Dirbc}) are obtained from (\ref%
{eigfunccirc}) by the replacement
\begin{equation}
J_{qn}(\gamma r)\rightarrow g_{qn}(\gamma r,\gamma a)\equiv J_{qn}(\gamma
r)Y_{qn}(\gamma a)-J_{qn}(\gamma a)Y_{qn}(\gamma r).  \label{replace}
\end{equation}%
Now the spectrum for the quantum number $\gamma $ is continuous and
\begin{equation}
\beta _{\alpha }^{2}=\frac{(2\pi )^{2-D}\gamma }{\phi _{0}\omega \left[
J_{qn}^{2}(\gamma a)+Y_{qn}^{2}(\gamma a)\right] }.  \label{norcoefext}
\end{equation}

Substituting the corresponding eigenfunctions into the mode sum formula (\ref%
{vevWf}), the positive frequency Whightman function in the exterior region
is presented in the form
\begin{eqnarray}
\langle 0|\varphi (x)\varphi (x^{\prime })|0\rangle &=&\frac{1}{\phi _{0}}%
\int \frac{d^{N}{\mathbf{k}}}{(2\pi )^{N}}e^{i{\mathbf{k}}\Delta \mathbf{r}%
_{\parallel }}\sum_{n=1}^{\infty }\sin (qn\phi )\sin (qn\phi ^{\prime })
\notag \\
&&\times \int_{0}^{\infty }d\gamma \frac{\gamma g_{qn}(\gamma r,\gamma
a)g_{qn}(\gamma r^{\prime },\gamma a)}{J_{qn}^{2}(\gamma
a)+Y_{qn}^{2}(\gamma a)}\frac{e^{-i\Delta t\sqrt{\gamma ^{2}+k_{m}^{2}}}}{%
\sqrt{\gamma ^{2}+k_{m}^{2}}}.  \label{Wfext0}
\end{eqnarray}%
To find the part in the Wightman function induced by the presence of the
cylindrical shell, we subtract from (\ref{Wfext0}) the corresponding
function for the wedge without a cylindrical shell, given by (\ref{Wf00}).
This allows to present the Wightman function in the form (\ref{Wf2}) with
the cylindrical shell induced part
\begin{eqnarray}
\langle \varphi (x)\varphi (x^{\prime })\rangle _{\mathrm{cyl}} &=&-\frac{2}{%
\pi \phi _{0}}\int \frac{d^{N}\mathbf{k}}{(2\pi )^{N}}e^{i\mathbf{k}\Delta
\mathbf{r}_{\parallel }}\int_{k}^{\infty }dz\frac{z\cosh (\Delta t\sqrt{%
z^{2}-k^{2}})}{\sqrt{z^{2}-k^{2}}}  \notag \\
&&\times \sum_{n=1}^{\infty }\sin (qn\phi )\sin (qn\phi ^{\prime
})K_{qn}(zr)K_{qn}(zr^{\prime })\frac{I_{qn}(za)}{K_{qn}(za)}.
\label{Wfa0ext}
\end{eqnarray}%
As we see, the expressions for the Wightman functions in the interior and
exterior regions are related by the interchange $I_{qn}\rightleftarrows
K_{qn}$ of the modified Bessel functions.

\section{VEVs inside a wedge without a cylindrical boundary}

\label{sec:Wedgecylabs}

In this section we consider the geometry of a wedge without a cylindrical
boundary. For integer values of $q$, after the explicit summation over $n$,
the Wightman function is presented in the form%
\begin{equation}
\langle 0_{w}|\varphi (x)\varphi (x^{\prime })|0_{w}\rangle =\frac{%
m^{(D-1)/2}}{(2\pi )^{(D+1)/2}}\sum_{j=1}^{2}(-1)^{j+1}\sum_{l=0}^{q-1}\frac{%
K_{(D-1)/2}(m\sqrt{u_{l}^{(j)2}+|\Delta \mathbf{r}_{\parallel }|^{2}-(\Delta
t)^{2}})}{\left[ u_{l}^{(j)2}+|\Delta \mathbf{r}_{\parallel }|^{2}-(\Delta
t)^{2}\right] ^{(D-1)/4}},  \label{Wf02}
\end{equation}%
where $u_{l}^{(j)}=\{r^{2}+r^{\prime 2}-2rr^{\prime }\cos [2\pi l/q+\phi
+(-1)^{j}\phi ^{\prime }]\}^{1/2}$. Note that the Wightman function in the
Minkowski spacetime coincides with the term $j=1$, $l=0$ in formula (\ref%
{Wf02}).

Taking the coincidence limit $x^{\prime }\rightarrow x$, for the difference
of the VEVs of the field squared,%
\begin{equation}
\langle \varphi ^{2}\rangle _{\mathrm{ren}}^{(w)}=\langle 0_{w}|\varphi
^{2}(x)|0_{w}\rangle -\langle 0_{M}|\varphi ^{2}(x)|0_{M}\rangle ,
\label{phiren}
\end{equation}%
where $|0_{M}\rangle $ is the amplitude for the vacuum state in the
Minkowski spacetime without boundaries, we find%
\begin{equation}
\langle \varphi ^{2}\rangle _{\mathrm{ren}}^{(w)}=\frac{m^{D-1}}{(2\pi
)^{(D+1)/2}}\sum_{j=1}^{2}(-1)^{j+1}\sideset{}{'}{\sum}_{l=0}^{q-1}\frac{%
K_{(D-1)/2}(2mr\sin \phi _{l}^{(j)})}{(2mr\sin \phi _{l}^{(j)})^{(D-1)/2}}.
\label{phi2wmass}
\end{equation}%
In this formula, the prime means that the term $j=1$, $l=0$ has to be
omitted, and we use the notation%
\begin{equation}
\phi _{l}^{(j)}=\pi l/q+(1+(-1)^{j})\phi /2.  \label{nj}
\end{equation}

For a massless field, from (\ref{phi2wmass}) we find
\begin{equation}
\langle \varphi ^{2}\rangle _{\mathrm{ren}}^{(w)}=\frac{\Gamma \left( \frac{%
D-1}{2}\right) }{(4\pi )^{\frac{D+1}{2}}r^{D-1}}\sum_{j=1}^{2}%
\sideset{}{'}{\sum}_{l=0}^{q-1}\frac{(-1)^{j+1}}{\sin ^{D-1}\phi _{l}^{(j)}},
\label{phi2w}
\end{equation}%
Note that the terms in this formula with $j=2$, $l=0$ and $j=2$, $l=q-1$ are
the corresponding VEVs for the geometry of a single plate located at $\phi =0
$ and $\phi =\phi _{0}$, respectively. In the case $D=3$ for the
renormalized VEV of the field square one finds \cite{Deutsch}%
\begin{equation}
\langle \varphi ^{2}\rangle _{\mathrm{ren}}^{(w)}=\frac{q^{2}-1-3q^{2}\csc
^{2}\left( q\phi \right) }{48\pi ^{2}r^{2}}.  \label{phi2D3}
\end{equation}

Near the wedge boundaries $\phi =\phi _{m}$, $m=0,1$ ($\phi _{1}=0$) the
main contribution in (\ref{phi2w}) comes from the terms $j=2$, $l=0$ and $%
l=q-1$ for $m=0$ and $m=1$ respectively, and the renormalized VEV of the
field squared diverges with the leading behaviour $\langle \varphi
^{2}\rangle _{\mathrm{ren}}^{(w)}\varpropto |\phi -\phi _{m}|^{1-D}$. The
surface divergences in the VEVs of the local physical observables are well
known in quantum field theory with boundaries and result from the
idealization of the boundaries as perfectly smooth surfaces which are
perfect reflectors at all frequencies. These divergences are investigated in
detail for various types of fields and general shape of smooth boundary \cite%
{Deutsch,Kenn80}. Near the smooth boundary the leading divergence in the
field squared varies as $(D-1)$th power of the distance from the boundary.
It seems plausible that such effects as surface roughness, or the
microstructure of the boundary on small scales (the atomic nature of matter
for the case of the electromagnetic field \cite{Cand82}) can introduce a
physical cutoff needed to produce finite values of surface quantities.

Now we turn to the VEVs of the energy-momentum tensor. By making use of
formula (\ref{vevEMTWf}), for the non-zero components one obtains (no
summation over $i$)%
\begin{equation}
\langle T_{i}^{i}\rangle _{\mathrm{ren}}^{(w)}=-\frac{\Gamma \left( \frac{D+1%
}{2}\right) }{2^{D+2}\pi ^{\frac{D+1}{2}}r^{D+1}}\sum_{j=1}^{2}%
\sideset{}{'}{\sum}_{l=0}^{q-1}\frac{(-1)^{j+1}f_{jl}^{(i)}}{\sin ^{D+1}\phi
_{l}^{(j)}},\;\langle T_{2}^{1}\rangle _{\mathrm{ren}}^{(w)}=\frac{D(\xi
_{c}-\xi )\Gamma \left( \frac{D+1}{2}\right) }{2^{D}\pi ^{\frac{D+1}{2}}r^{D}%
}\sum_{l=0}^{q-1}\frac{\cos \phi _{l}^{(2)}}{\sin ^{D}\phi _{l}^{(2)}},
\label{Tiiw21}
\end{equation}%
where $i=0,1,\ldots ,D$, and we use the following notations%
\begin{eqnarray}
f_{jl}^{(i)} &=&1+\left( 4\xi -1\right) \left[ (D-1)\delta _{j1}\sin
^{2}\phi _{l}^{(j)}+D\delta _{j2}\right] ,\quad i=0,3,\ldots ,D,  \label{fji}
\\
f_{jl}^{(1)} &=&f_{jl}^{(0)}-4D(\xi -\xi _{c})\sin ^{2}\phi
_{l}^{(j)},\;f_{jl}^{(2)}=D\left[ 4\sin ^{2}\phi _{l}^{(j)}\left( \xi -\xi
_{c}\delta _{j2}\right) -\delta _{j1}\right] .  \label{fji2}
\end{eqnarray}%
In the case $\phi _{0}=\pi /2$ and for minimally and conformally coupled
scalar fields, it can be checked that from formulae (\ref{Tiiw21}), after
the transformation from cylindrical coordinates to the cartesian ones, as a
special case we obtain the result derived in \cite{Acto96}. For a
conformally coupled scalar field $f_{2l}^{(i)}=0$ and from (\ref{Tiiw21})
one finds%
\begin{equation}
\langle T_{i}^{k}\rangle _{\mathrm{ren}}^{(w)}=-\frac{\Gamma \left( \frac{D+1%
}{2}\right) }{2^{D+2}\pi ^{\frac{D+1}{2}}r^{D+1}}\sum_{l=1}^{q-1}\frac{%
D-(D-1)\sin ^{2}(\pi l/q)}{D\sin ^{D+1}(\pi l/q)}\mathrm{diag}%
(1,1,-D,1,\ldots ,1).  \label{TikwConf}
\end{equation}%
In this case the vacuum energy-momentum tensor does not depend on the
angular coordinate. For a non-conformally coupled field the VEVs (\ref%
{Tiiw21}) diverge on the boundaries $\phi =\phi _{m}$ and for points away
from the edge $r=0$, these divergences are the same as those for the
geometry of a single plate.

In the most important case $D=3$, for the components of the renormalized
energy-momentum tensor we find%
\begin{eqnarray}
\langle T_{0}^{0}\rangle _{\mathrm{ren}}^{(w)} &=&\langle T_{3}^{3}\rangle _{%
\mathrm{ren}}^{(w)}=\frac{1}{32\pi ^{2}r^{4}}\left\{ \frac{1-q^{4}}{45}+%
\frac{8}{3}\left( 1-q^{2}\right) (\xi -\xi _{c})\right.  \notag \\
&&\left. +12\frac{(\xi -\xi _{c})q^{2}}{\sin ^{2}(q\phi )}\left[ \frac{q^{2}%
}{\sin ^{2}(q\phi )}-\frac{2}{3}q^{2}+\frac{2}{3}\right] \right\} ,
\label{TikD3} \\
\langle T_{1}^{1}\rangle _{\mathrm{ren}}^{(w)} &=&\frac{1}{32\pi ^{2}r^{4}}%
\left\{ \frac{1-q^{4}}{45}-\frac{4}{3}(1-q^{2})(\xi -\xi _{c})\right.  \notag
\\
&&\left. +12\frac{(\xi -\xi _{c})q^{2}}{\sin ^{2}\left( q\phi \right) }\left[
\frac{q^{2}}{\sin ^{2}\left( q\phi \right) }-\frac{2}{3}q^{2}-\frac{1}{3}%
\right] \right\} ,  \label{TikD311} \\
\langle T_{2}^{1}\rangle _{\mathrm{ren}}^{(w)} &=&-\frac{3(\xi -\xi _{c})}{%
8\pi ^{2}r^{3}}\frac{q^{3}\cos \left( q\phi \right) }{\sin ^{3}\left( q\phi
\right) },  \label{TikD321} \\
\langle T_{2}^{2}\rangle _{\mathrm{ren}}^{(w)} &=&\frac{1}{8\pi ^{2}r^{4}}%
\left[ \frac{q^{4}-1}{60}+(\xi -\xi _{c})\left( 1-q^{2}+\frac{3q^{2}}{\sin
^{2}\left( q\phi \right) }\right) \right] .  \label{TikD322}
\end{eqnarray}%
Though we have derived these formulae for integer values of the parameter $q$%
, by the analytic continuation they are valid for non-integer values of this
parameter as well. For a conformally coupled scalar field we obtain the
result previously derived in the literature \cite{jphy,Deutsch}. The
corresponding vacuum forces acting on the wedge sides are determined by the
effective pressure $-\langle T_{2}^{2}\rangle _{\mathrm{ren}}^{(w)}$. These
forces are attractive for the wedge with $q>1$ and are repulsive for $q<1$.

\section{Field squared and the energy-momentum tensor}

\label{sec:Wedgecyl}

We now turn to the geometry of a wedge with additional cylindrical boundary
of radius $a$. Taking the coincidence limit $x^{\prime }\rightarrow x$ in
formula (\ref{Wf2}) for the Wightman function and integrating over $\mathbf{k%
}$, the VEV of the field squared is presented as the sum of two terms:%
\begin{equation}
\langle 0|\varphi ^{2}|0\rangle =\langle 0_{w}|\varphi ^{2}|0_{w}\rangle
+\langle \varphi ^{2}\rangle _{\mathrm{cyl}},  \label{phi2a}
\end{equation}%
where the part induced by the cylindrical boundary is given by the formula%
\begin{equation}
\langle \varphi ^{2}\rangle _{\mathrm{cyl}}=-\frac{2^{3-D}\pi ^{\frac{1-D}{2}%
}}{\Gamma \left( \frac{D-1}{2}\right) \phi _{0}}\sum_{n=1}^{\infty }\sin
^{2}(qn\phi )\int_{m}^{\infty }dz\,z\left( z^{2}-m^{2}\right) ^{\frac{D-3}{2}%
}\frac{K_{qn}(az)}{I_{qn}(az)}I_{qn}^{2}(rz).  \label{phi2a1}
\end{equation}%
Note that this part vanishes at the wedge sides $\phi =\phi _{m}$, $%
0\leqslant r<a$. Near the edge $r=0$ the main contribution into $\langle
\varphi ^{2}\rangle _{\mathrm{cyl}}$ comes from the term $n=1$ and $\langle
\varphi ^{2}\rangle _{\mathrm{cyl}}$ behaves like $r^{2q}$. The part $%
\langle \varphi ^{2}\rangle _{\mathrm{cyl}}$ diverges on the cylindrical
surface $r=a$. Near this surface the main contribution into (\ref{phi2a1})
comes from large values $n$ and for $|\phi -\phi _{m}|\gg 1-r/a$ the leading
behavior is the same as that for a cylindrical surface of radius $a$.

Similarly, the VEV of the energy-momentum tensor for the situation when the
cylindrical boundary is present is written in the form%
\begin{equation}
\langle 0|T_{ik}|0\rangle =\langle 0_{w}|T_{ik}|0_{w}\rangle +\langle
T_{ik}\rangle _{\mathrm{cyl}},  \label{Tika}
\end{equation}%
where $\langle T_{ik}\rangle _{\mathrm{cyl}}$ is induced by the cylindrical
boundary. This term is obtained from the corresponding part in the Wightman
function, $\langle \varphi (x)\varphi (x^{\prime })\rangle _{\mathrm{cyl}}$,
by using formula (\ref{vevEMTWf}). For points away from the cylindrical
surface this limit gives a finite result. For the corresponding components
of the energy-momentum tensor one obtains (no summation over $i$)%
\begin{eqnarray}
\langle T_{i}^{i}\rangle _{\mathrm{cyl}} &=&\frac{(4\pi )^{-\frac{D-1}{2}}}{%
\Gamma \left( \frac{D-1}{2}\right) \phi _{0}}\sum_{n=1}^{\infty
}\int_{m}^{\infty }dz\,z^{3}\left( z^{2}-m^{2}\right) ^{\frac{D-3}{2}}\frac{%
K_{qn}(az)}{I_{qn}(az)}  \notag \\
&&\times \left\{ a_{i,qn}^{(+)}[I_{qn}(rz)]-a_{i,qn}^{(-)}[I_{qn}(rz)]\cos
(2qn\phi )\right\} ,  \label{Tiia} \\
\langle T_{2}^{1}\rangle _{\mathrm{cyl}} &=&\frac{2(4\pi )^{-\frac{D-1}{2}}}{%
\Gamma \left( \frac{D-1}{2}\right) \phi _{0}}\sum_{n=1}^{\infty }qn\sin
(2qn\phi )\int_{m}^{\infty }dz\,z^{2}(z^{2}-m^{2})^{\frac{D-3}{2}}\frac{%
K_{qn}(az)}{I_{qn}(az)}  \notag \\
&&\times I_{qn}(rz)\left[ \frac{2\xi }{rz}I_{qn}(rz)+(1-4\xi )I_{qn}^{\prime
}(rz)\right] ,  \label{Tiia21}
\end{eqnarray}%
with the notations%
\begin{eqnarray}
a_{i,l}^{(\pm )}[g(y)] &=&(4\xi -1)\left[ g^{\prime 2}(y)+\left( 1\pm
l^{2}/y^{2}\right) g^{2}(y)\right] +2g^{2}(y)\frac{1-m^{2}r^{2}/y^{2}}{D-1},
\notag \\
a_{1,l}^{(\pm )}[g(y)] &=&g^{\prime 2}(y)+(4\xi /y)g(y)g^{\prime
}(y)-g^{2}(y)\left\{ 1\pm \left[ 1-4\xi (1\mp 1)\right] l^{2}/y^{2}\right\} ,
\label{ajpm1} \\
a_{2,l}^{(\pm )}[g(y)] &=&\left( 4\xi -1\right) \left[ g^{\prime
2}(y)+g^{2}(y)\right] -(4\xi /y)g(y)g^{\prime }(y)+g^{2}(y)\left( 4\xi \pm
1\right) l^{2}/y^{2},  \notag
\end{eqnarray}%
for a given function $g(y)$, $i=0,3,\ldots ,D$. In accordance with the
problem symmetry, the expressions for the diagonal components are invariant
under the replacement $\phi \rightarrow \phi _{0}-\phi $, and the
off-diagonal component $\langle T_{2}^{1}\rangle _{\mathrm{cyl}}$ changes
the sign under this replacement. Note that the latter vanishes on the wedge
sides $\phi =\phi _{m}$, $0\leqslant r<a$ and for $\phi =\phi _{0}/2$. On
the wedge sides for the diagonal components of the energy-momentum tensor we
obtain (no summation over $i$)%
\begin{equation}
\langle T_{i}^{i}\rangle _{\mathrm{cyl},\phi =\phi _{m}}=\frac{2^{2-D}\pi ^{%
\frac{5-D}{2}}A_{i}}{\Gamma \left( \frac{D-1}{2}\right) r^{2}\phi _{0}^{3}}%
\sum_{n=1}^{\infty }n^{2}\int_{m}^{\infty }dz\,z\left( z^{2}-m^{2}\right) ^{%
\frac{D-3}{2}}\frac{K_{qn}(az)}{I_{qn}(az)}I_{qn}^{2}(rz),  \label{Tiionphim}
\end{equation}%
where $A_{i}=4\xi -1$, $i=0,1,3,\ldots ,D$, $A_{2}=1$. In particular, the
additional vacuum effective pressure in the direction perpendicular to the
wedge sides, $p_{a}=-\langle T_{2}^{2}\rangle _{\mathrm{cyl},\phi =\phi _{m}}
$, does not depend on the curvature coupling parameter and is negative for
all values $0<r<a$. This means that the vacuum forces acting on the wedge
sides due to the presence of the cylindrical boundary are attractive. The
corresponding vacuum stresses in the directions parallel to the wedge sides
are isotropic and the energy density is negative for both minimally and
conformally coupled scalars.

For $0<r<a$ the cylindrical parts (\ref{Tiia}) and (\ref{Tiia21}) are finite
for all values $0\leqslant \phi \leqslant \phi _{0}$, including the wedge
sides. The divergences on these sides are included in the first term on the
right-hand side of (\ref{Tika}) corresponding to the case without
cylindrical boundary. Near the edge $r=0$ the main contribution into the
boundary parts comes from the summand with $n=1$ and one has $\langle
T_{i}^{i}\rangle _{\mathrm{cyl}}\propto r^{2q-2}$, $\langle T_{2}^{1}\rangle
_{\mathrm{cyl}}\propto r^{2q-1}$. The boundary part $\left\langle
T_{i}^{k}\right\rangle _{\mathrm{cyl}}$ diverges on the cylindrical surface $%
r=a$. Expanding over $a-r$, on the wedge sides for the diagonal components
one finds
\begin{equation}
\langle T_{i}^{i}\rangle _{\mathrm{cyl},\phi =\phi _{m}}\approx \frac{%
A_{i}\Gamma \left( \frac{D+1}{2}\right) }{2(4\pi )^{\frac{D+1}{2}}(a-r)^{D+1}%
},\quad r\rightarrow a,  \label{T00asra1}
\end{equation}%
where the coefficients $A_{i}$ are defined in the paragraph after formula (%
\ref{Tiionphim}). It can be seen that for the off-diagonal component to the
leading order one has $\langle T_{2}^{1}\rangle _{a}\varpropto (a-r)^{-D}$.
For angles $0<\phi <\phi _{0}$, and for $|\phi -\phi _{m}|\gg 1-r/a$, the
leading divergence coincides with the corresponding one for a cylindrical
surface of the radius $a$.

Taking the coincidence limit of the arguments, from formula (\ref{Wfa0ext})
we obtain the VEV of the field squared in the region $r>a$:
\begin{equation}
\langle \varphi ^{2}\rangle _{\mathrm{cyl}}=-\frac{2^{3-D}\pi ^{\frac{1-D}{2}%
}}{\Gamma \left( \frac{D-1}{2}\right) \phi _{0}}\sum_{n=1}^{\infty }\sin
^{2}(qn\phi )\int_{m}^{\infty }dz\,z\left( z^{2}-m^{2}\right) ^{\frac{D-3}{2}%
}\frac{I_{qn}(az)}{K_{qn}(az)}K_{qn}^{2}(rz).  \label{phi2a1ext}
\end{equation}%
As for the interior region, the VEV (\ref{phi2a1ext}) diverges on the
cylindrical surface. For large distances from the cylindrical surface, $r\gg
a$, and for a massless field the main contribution comes from the $n=1$ term
and to the leading order one finds $\langle \varphi ^{2}\rangle _{\mathrm{cyl%
}}\varpropto (a/r)^{D-1+2q}$. For a massive field and for $mr\gg 1$ the part
$\langle \varphi ^{2}\rangle _{\mathrm{cyl}}$ is exponentially suppressed.

For the part in the vacuum energy-momentum tensor induced by the cylindrical
surface in the region $r>a$, from (\ref{vevEMTWf}), (\ref{Wfa0ext}), (\ref%
{phi2a1ext}) one has the following formulae
\begin{eqnarray}
\langle T_{i}^{i}\rangle _{\mathrm{cyl}} &=&\frac{(4\pi )^{-\frac{D-1}{2}}}{%
\Gamma \left( \frac{D-1}{2}\right) \phi _{0}}\sum_{n=1}^{\infty
}\int_{m}^{\infty }dz\,z^{3}\left( z^{2}-m^{2}\right) ^{\frac{D-3}{2}}\frac{%
I_{qn}(az)}{K_{qn}(az)}  \notag \\
&&\times \left\{ a_{i,qn}^{(+)}[K_{qn}(rz)]-a_{i,qn}^{(-)}[K_{qn}(rz)]\cos
(2qn\phi )\right\} ,  \label{Tiiaext} \\
\langle T_{2}^{1}\rangle _{\mathrm{cyl}} &=&\frac{2(4\pi )^{-\frac{D-1}{2}}}{%
\Gamma \left( \frac{D-1}{2}\right) \phi _{0}}\sum_{n=1}^{\infty }qn\sin
(2qn\phi )\int_{m}^{\infty }dz\,\,z^{2}(z^{2}-m^{2})^{\frac{D-3}{2}}\frac{%
I_{qn}(az)}{K_{qn}(az)}  \notag \\
&&\times K_{qn}(rz)\left[ \frac{2\xi }{rz}K_{qn}(rz)+(1-4\xi )K_{qn}^{\prime
}(rz)\right] ,  \label{Tiia21ext}
\end{eqnarray}%
with the functions $a_{i,qn}^{(\pm )}[g(y)]$ defined by (\ref{ajpm1}). In
the way similar to that used above for the VEV of the field square, it can
be seen that at large distances from the cylindrical surface, $r\gg a$, the
main contribution comes from the term with $n=1$ and for a massless field
the components of the induced energy-momentum tensor behave as $\langle
T_{i}^{i}\rangle _{\mathrm{cyl}}\varpropto (a/r)^{D+1+2q}$, $\langle
T_{2}^{1}\rangle _{\mathrm{cyl}}\varpropto (a/r)^{D+2q}$. As for the
interior region, the vacuum forces acting on the wedge sides due to the
presence of the cylindrical shell are attractive and the corresponding
energy density is negative for both minimally and conformally coupled
scalars.

In the limit $\phi _{0}\rightarrow 0$, $r,a\rightarrow \infty $, assuming
that $a-r$ and $a\phi _{0}\equiv b$ are fixed, from the results given above
we obtain the vacuum densities for the geometry of two parallel plates
separated by a distance $b$, perpendicularly intersected by the third plate.
The vacuum expectation values of the energy-momentum tensor for this
geometry of boundaries are investigated in \cite{Acto96} for special cases
of minimally and conformally coupled massless scalar fields.

\section{VEVs for the electromagnetic fields}

\label{sec:Inter}

\subsection{Interior region}

In this section we consider a wedge with a coaxial cylindrical boundary
assuming that all boundaries are perfectly conducting. For this geometry
there are two different types of the eigenfunctions corresponding to the
transverse magnetic (TM, $\lambda =0$) and transverse electric (TE, $\lambda
=1$) waves. In the Coulomb gauge, the vector potentials for these modes are
given by the formulae%
\begin{equation}
\mathbf{A}_{\alpha }=\beta _{\alpha }\left\{
\begin{array}{cc}
(1/i\omega )\left( \gamma ^{2}\mathbf{e}_{3}+ik\nabla _{t}\right)
J_{qn}(\gamma r)\sin (qn\phi )\exp \left[ i\left( kz-\omega t\right) \right]
, & \lambda =0 \\
-\mathbf{e}_{3}\times \nabla _{t}\left\{ J_{qn}(\gamma r)\cos (qn\phi )\exp %
\left[ i\left( kz-\omega t\right) \right] \right\} , & \lambda =1%
\end{array}%
\right. ,  \label{Aalpha}
\end{equation}%
where $\mathbf{e}_{3}$ is the unit vector along the axis of the wedge, $%
\nabla _{t}$ is the part of the nabla operator transverse to this axis, and $%
\omega ^{2}=\gamma ^{2}+k^{2}$. In Eq. (\ref{Aalpha}), $n=1,2,\ldots $ for $%
\lambda =0$ and $n=0,1,2,\ldots $ for $\lambda =1$. From the normalization
condition one finds%
\begin{equation}
\beta _{\alpha }^{2}=\frac{4qT_{qn}(\gamma a)}{\pi \omega a\gamma }\delta
_{n},\;\delta _{n}=\left\{
\begin{array}{cc}
1/2, & n=0 \\
1, & n\neq 0%
\end{array}%
\right. ,  \label{betalfel}
\end{equation}%
where we have introduced the notation $T_{\nu }(x)=x\left[ J_{\nu
}^{^{\prime }2}(x)+(1-\nu ^{2}/x^{2})J_{\nu }^{2}(x)\right] ^{-1}$.
Eigenfunctions (\ref{Aalpha}) satisfy the standard boundary conditions on
the wedge sides. From the boundary conditions on the cylindrical shell it
follows that the eigenvalues for $\gamma $ are roots of the equation
\begin{equation}
J_{qn}^{(\lambda )}(\gamma a)=0,\quad \lambda =0,1,  \label{modes1}
\end{equation}%
where $J_{\nu }^{(0)}(x)=J_{\nu }(x)$ and $J_{\nu }^{(1)}(x)=J_{\nu
}^{\prime }(x)$. We will denote the corresponding eigenmodes by $\gamma
a=\lambda _{n,j}^{(\lambda )}$, $j=1,2,\ldots $.

First we consider the VEVs of the squares of the electric and magnetic
fields inside the shell. Substituting the eigenfunctions (\ref{Aalpha}) into
the corresponding mode-sum formula, we find%
\begin{equation}
\langle 0|F^{2}|0\rangle =\frac{4q}{\pi a^{3}}\sideset{}{'}{\sum}%
_{m=0}^{\infty }\int_{-\infty }^{+\infty }dk\sum_{\lambda
=0,1}\sum_{n=1}^{\infty }\frac{\lambda _{n,j}^{(\lambda )3}T_{qm}(\lambda
_{n,j}^{(\lambda )})}{\sqrt{\lambda _{n,j}^{(\lambda )2}+k^{2}a^{2}}}%
g^{(\eta _{F\lambda })}[\Phi _{qn}^{(\lambda )}(\phi ),J_{qn}(\lambda
_{n,j}^{(\lambda )}r/a)],  \label{F2}
\end{equation}%
where $F=E,B$ with $\eta _{E\lambda }=\lambda $, $\eta _{B\lambda
}=1-\lambda $, and the prime in the summation over $n$ means that the term $%
n=0$ should be halved. In formula (\ref{F2}) we have introduced the notations%
\begin{eqnarray}
g^{(0)}[\Phi (\phi ),f(x)] &=&(k^{2}r^{2}/x^{2})\left[ \Phi ^{2}(\phi
)f^{\prime 2}(x)+\Phi ^{\prime 2}(\phi )f^{2}(x)/x^{2}\right] +\Phi
^{2}(\phi )f^{2}(x),  \notag \\
g^{(1)}[\Phi (\phi ),f(x)] &=&(1+k^{2}r^{2}/x^{2})\left[ \Phi ^{2}(\phi
)f^{\prime 2}(x)+\Phi ^{\prime 2}(\phi )f^{2}(x)/x^{2}\right] ,
\label{gnulam2}
\end{eqnarray}%
and%
\begin{equation}
\Phi _{\nu }^{(0)}(\phi )=\sin (\nu \phi ),\;\Phi _{\nu }^{(1)}(\phi )=\cos
(\nu \phi ).  \label{Philam}
\end{equation}%
The expressions (\ref{F2}) corresponding to the electric and magnetic fields
are divergent. They may be regularized introducing a cutoff function $\psi
_{\mu }(\omega )$ with the cutting parameter $\mu $ which makes the
divergent expressions finite and satisfies the condition $\psi _{\mu
}(\omega )\rightarrow 1$ for $\mu \rightarrow 0$. After the renormalization
the cutoff function is removed by taking the limit $\mu \rightarrow 0$.

In order to further simplify the VEVs, we apply to the series over $n$ the
summation formula (\ref{sumform1AP}) for the modes with $\lambda =0$ and the
similar formula from \cite{Sahrev} for the modes with $\lambda =1$. As it
can be seen, for points away from the shell the contribution to the VEVs
coming from the second integral terms on the right-hand sides of these
formulae are finite in the limit $\mu \rightarrow 0$ and, hence, the cutoff
function in these terms can be safely removed. As a result the VEVs are
written in the form%
\begin{equation}
\langle 0|F^{2}|0\rangle =\langle 0_{w}|F^{2}|0_{w}\rangle +\left\langle
F^{2}\right\rangle _{\mathrm{cyl}},  \label{F21}
\end{equation}%
where%
\begin{eqnarray}
\langle 0_{w}|F^{2}|0_{w}\rangle  &=&\frac{q}{\pi }\sideset{}{'}{\sum}%
_{n=0}^{\infty }\int_{-\infty }^{+\infty }dk\int_{0}^{\infty }d\gamma \,%
\frac{\gamma ^{3}\psi _{\mu }(\omega )}{\sqrt{\gamma ^{2}+k^{2}}}\left\{
\left( 1+\frac{2k^{2}}{\gamma ^{2}}\right) \left[ J_{qn}^{\prime 2}(\gamma
r)+\frac{q^{2}n^{2}}{\gamma ^{2}r^{2}}J_{qn}^{2}(\gamma r)\right] \right.
\notag \\
&&\left. +J_{qn}^{2}(\gamma r)-(-1)^{\eta _{F1}}\cos (2qn\phi )\left[
J_{qn}^{\prime 2}(\gamma r)-\left( 1+\frac{q^{2}n^{2}}{\gamma ^{2}r^{2}}%
\right) J_{qn}^{2}(\gamma r)\right] \right\} ,  \label{F2s}
\end{eqnarray}%
and%
\begin{equation}
\langle F^{2}\rangle _{\mathrm{cyl}}=\frac{2q}{\pi }\sideset{}{'}{\sum}%
_{n=0}^{\infty }\sum_{\lambda =0,1}\int_{0}^{\infty }dx\,x^{3}\frac{%
K_{qn}^{(\lambda )}(xa)}{I_{qn}^{(\lambda )}(xa)}G^{(\eta _{F\lambda
})}[\Phi _{qn}^{(\lambda )}(\phi ),I_{qn}(xr)].  \label{F2b0}
\end{equation}%
In formula (\ref{F2b0}) we have introduced the notations%
\begin{eqnarray}
G^{(0)}[\Phi (\phi ),f(x)] &=&\Phi ^{2}(\phi )f^{\prime 2}(x)+\Phi ^{\prime
2}(\phi )f^{2}(x)/x^{2}+2\Phi ^{2}(\phi )f^{2}(x),  \notag \\
G^{(1)}[\Phi (\phi ),f(x)] &=&-\Phi ^{2}(\phi )f^{\prime 2}(x)-\Phi ^{\prime
2}(\phi )f^{2}(x)/x^{2}.  \label{Gnujtilde1}
\end{eqnarray}%
The second term on the right-hand side of Eq. (\ref{F21}) vanishes in the
limit $a\rightarrow \infty $ and the first one does not depend on $a$. Thus,
we can conclude that the term $\langle 0_{w}|F^{2}|0_{w}\rangle $
corresponds to the part in the VEVs\ when the cylindrical shell is absent.

First, let us concentrate on the part corresponding to the wedge without a
cylindrical shell. In (\ref{F2s}) the part which does not depend on the
angular coordinate $\phi $ is the same as in the corresponding problem of
the cosmic string geometry with the angle deficit $2\pi -\phi _{0}$ (see
\cite{Beze06}), which we will denote by $\langle 0_{\mathrm{s}}|F^{2}|0_{%
\mathrm{s}}\rangle $. For this part we have
\begin{equation}
\langle 0_{\mathrm{s}}|F^{2}|0_{\mathrm{s}}\rangle =\langle 0_{\mathrm{M}%
}|F^{2}|0_{\mathrm{M}}\rangle -\frac{(q^{2}-1)(q^{2}+11)}{180\pi r^{4}},
\label{F2sn}
\end{equation}%
where $\langle 0_{\mathrm{M}}|F^{2}|0_{\mathrm{M}}\rangle $ is the VEV in
the Minkowski spacetime without boundaries and in the last expression we
have removed the cutoff. To evaluate the part in (\ref{F2s}) which depends
on $\phi $, we firstly consider the case when the parameter $q$ is an
integer. In this case, the summation over $n$ can be done explicitly and the
integrals are evaluated by introducing polar coordinates in the $(k,\gamma )$%
-plane. As a result, for the renormalised VEVs\ of the field squared in the
geometry of a wedge without a cylindrical boundary we find%
\begin{equation}
\langle F^{2}\rangle _{\mathrm{ren}}^{(w)}=-\frac{(q^{2}-1)(q^{2}+11)}{%
180\pi r^{4}}-\frac{(-1)^{\eta _{F1}}q^{2}}{2\pi r^{4}\sin ^{2}(q\phi )}%
\left[ 1-q^{2}+\frac{3q^{2}}{2\sin ^{2}(q\phi )}\right] ,  \label{F2wren}
\end{equation}%
with $\eta _{E1}=1$ and $\eta _{B1}=0$. Though we have derived this formula
for integer values of the parameter $q$, by the analytic continuation it is
valid for non-integer values of this parameter as well. The expression on
the right of formula (\ref{F2wren}) is invariant under the replacement $\phi
\rightarrow \phi _{0}-\phi $ and, as we could expect, the VEVs are symmetric
with respect to the half-plane $\phi =\phi _{0}/2$. Formula (\ref{F2wren})
for $F=E$ was derived in Ref. \cite{Brev98} within the framework of
Schwinger's source theory.

Now, we turn to the investigation of the parts in the VEVs of the field
squared induced by the cylindrical boundary and given by formula (\ref{F2b0}%
). These parts are symmetric with respect to the half-plane $\phi =\phi
_{0}/2$. The expression in the right-hand side of (\ref{F2b0}) is finite for
$0<r<a$ including the points on the wedge sides, and diverges on the shell.
To find the leading term in the corresponding asymptotic expansion, we note
that near the shell the main contribution comes from large values of $n$. By
using the uniform asymptotic expansions of the modified Bessel functions for
large values of the order, up to the leading order, for the points $a-r\ll
a|\sin \phi |,a|\sin (\phi _{0}-\phi )|$ we find $\langle F^{2}\rangle _{%
\mathrm{cyl}}\approx -3(-1)^{\eta _{F1}}/[4\pi (a-r)^{4}]$. These surface
divergences originate in the unphysical nature of perfect conductor boundary
conditions. In reality the expectation values will attain a limiting value
on the conductor surface, which will depend on the molecular details of the
conductor. From the formulae given above it follows that the main
contribution to $\langle F^{2}\rangle _{\mathrm{cyl}}$ are due to the
frequencies $\omega \lesssim (a-r)^{-1}$. Hence, we expect that formula (\ref%
{F2b0}) is valid for real conductors up to distances $r$ for which $%
(a-r)^{-1}\ll \omega _{0}$, with $\omega _{0}$ being the characteristic
frequency, such that for $\omega >\omega _{0}$ the conditions for perfect
conductivity fail.

Near the edge $r=0$, assuming that $r/a\ll 1$, the asymptotic behavior of
the part induced in the VEVs of the field squared by the cylindrical shell
depends on the parameter $q$. For $q>1+\eta _{F1}$, the dominant
contribution comes from the lowest mode $n=0$ and to the leading order one
has $\langle F^{2}\rangle _{\mathrm{cyl}}\varpropto r^{2\eta _{F1}}$. In
this case the quantity $\langle B^{2}\rangle _{\mathrm{cyl}}$ takes a finite
limiting value on the edge $r=0$, whereas $\langle E^{2}\rangle _{\mathrm{cyl%
}}$ vanishes as $r^{2}$. For $q<1+\eta _{F1}$ the main contribution comes
from the mode with $n=1$ and the shell-induced parts diverge on the edge $r=0
$ with $\langle F^{2}\rangle _{\mathrm{cyl}}\varpropto r^{2(q-1)}$. In
accordance with (\ref{F2wren}), near the edge $r=0$ the total VEV is
dominated by the part coming from the wedge without the cylindrical shell.

\subsection{Exterior region}

In the exterior region (region II in figure \ref{fig1}), the corresponding
eigenfunctions for the vector potential are obtained from formulae (\ref%
{Aalpha}) by the replacement%
\begin{equation}
J_{qn}(\gamma r)\rightarrow g_{qn}^{(\lambda )}(\gamma a,\gamma
r)=J_{qn}(\gamma r)Y_{qn}^{(\lambda )}(\gamma a)-Y_{qn}(\gamma
r)J_{qn}^{(\lambda )}(\gamma a),  \label{extreplace}
\end{equation}%
where, as before, $\lambda =0,1$ correspond to the waves of the electric and
magnetic types, respectively. The eigenvalues for $\gamma $ are continuous
and
\begin{equation}
\beta _{\alpha }^{-2}=(8\pi /q)\delta _{n}\gamma \omega \left[
J_{qn}^{(\lambda )2}(\gamma a)+Y_{qn}^{(\lambda )2}(\gamma a)\right] .
\label{betalfext}
\end{equation}%
Substituting the eigenfunctions into the corresponding mode-sum formula, for
the VEV of the field squared one finds%
\begin{equation}
\langle 0|F^{2}|0\rangle =\frac{2q}{\pi }\sideset{}{'}{\sum}_{n=0}^{\infty
}\int_{-\infty }^{+\infty }dk\int_{0}^{\infty }d\gamma \sum_{\lambda =0,1}%
\frac{\gamma ^{3}}{\sqrt{k^{2}+\gamma ^{2}}}\frac{g^{(\eta _{F\lambda
})}[\Phi _{qn}^{(\lambda )}(\phi ),g_{qn}^{(\lambda )}(\gamma a,\gamma r)]}{%
J_{qn}^{(\lambda )2}(\gamma a)+Y_{qn}^{(\lambda )2}(\gamma a)},
\label{F2ext}
\end{equation}%
where the functions $g^{(\eta _{F\lambda })}[\Phi (\phi ),f(x)]$ are defined
by relations (\ref{gnulam2}) with $f(x)=g_{qn}^{(\lambda )}(\gamma a,x)$. To
extract from this VEV the part induced by the cylindrical shell, we subtract
from the right-hand side the corresponding expression for the wedge without
the cylindrical boundary. As a result, the VEV of the field squared is
written in the form (\ref{F21}), where the part induced by the cylindrical
shell is given by the formula
\begin{equation}
\langle F^{2}\rangle _{\mathrm{cyl}}=\frac{2q}{\pi }\sideset{}{'}{\sum}%
_{n=0}^{\infty }\sum_{\lambda =0,1}\int_{0}^{\infty }dx\,x^{3}\frac{%
I_{qn}^{(\lambda )}(xa)}{K_{qn}^{(\lambda )}(xa)}G^{(\eta _{F\lambda
})}[\Phi _{qn}^{(\lambda )}(\phi ),K_{qn}(xr)].  \label{F2bext}
\end{equation}%
In this formula the functions $G^{(\eta _{F\lambda })}\left[ \Phi (\phi
),f(x)\right] $ are defined by expressions (\ref{Gnujtilde1}). Comparing
this result with formula (\ref{F2b0}), we see that the expressions for the
shell-induced parts in the interior and exterior regions are related by the
interchange $I_{qn}\rightleftarrows K_{qn}$.

The VEV (\ref{F2bext}) diverges on the cylindrical shell with the leading
term being the same as that for the interior region. At large distances from
the cylindrical shell we introduce a new integration variable $y=xr$ and
expand the integrand over $a/r$. For $q>1$ the main contribution comes from
the lowest mode $n=0$ and up to the leading order we have
\begin{equation}
\langle E^{2}\rangle _{\mathrm{cyl}}\approx \frac{4q\left( a/r\right) ^{2}}{%
5\pi r^{4}},\;\langle B^{2}\rangle _{\mathrm{cyl}}\approx -\frac{28q\left(
a/r\right) ^{2}}{15\pi r^{4}}.  \label{B2far}
\end{equation}%
For $q<1$ the dominant contribution into the VEVs at large distances is due
to the mode $n=1$ with the leading term%
\begin{equation}
\langle F^{2}\rangle _{\mathrm{cyl}}\approx -\frac{4q^{2}(q+1)}{\pi r^{4}}%
\left( \frac{a}{r}\right) ^{2q}\left[ \frac{\cos (2q\phi )}{2q+3}+(-1)^{\eta
_{F1}}\frac{q+1}{2q+1}\right] .  \label{F2far}
\end{equation}%
For the case $q=1$ the contributions of the modes $n=0$ and $n=1$ are of the
same order and the corresponding leading terms are obtained by summing these
contributions. The latter are given by the right-hand sides of formulae (\ref%
{B2far}) and (\ref{F2far}). As we see, at large distances the part induced
by the cylindrical shell is suppressed with respect to the part
corresponding to the wedge without the shell by the factor $(a/r)^{2\beta }$
with $\beta =\min (1,q)$.

\section{Energy-momentum tensor for the electromagnetic field}

\label{sec:EMTint}

Now let us consider the VEV of the energy-momentum tensor in the region
inside the cylindrical shell. Substituting the eigenfunctions (\ref{Aalpha})
into the corresponding mode-sum formula, for the non-zero components we
obtain (no summation over $i$)%
\begin{eqnarray}
\langle 0|T_{i}^{i}|0\rangle  &=&\frac{q}{2\pi ^{2}a^{3}}\sideset{}{'}{\sum}%
_{n=0}^{\infty }\int_{-\infty }^{+\infty }dk\sum_{\lambda
=0,1}\sum_{j=1}^{\infty }\frac{\lambda _{n,j}^{(\lambda )3}T_{qn}(\lambda
_{n,j}^{(\lambda )})}{\sqrt{\lambda _{n,j}^{(\lambda )2}+k^{2}a^{2}}}%
f^{(i)}[\Phi _{qn}^{(\lambda )}(\phi ),J_{qn}(\lambda _{n,j}^{(\lambda
)}r/a)],  \label{Tik} \\
\langle 0|T_{2}^{1}|0\rangle  &=&\frac{-q^{2}}{4\pi ^{2}a}\frac{\partial }{%
\partial r}\sideset{}{'}{\sum}_{n=0}^{\infty }n\sin (2qn\phi )\int_{-\infty
}^{+\infty }dk\sum_{\lambda =0,1}(-1)^{\lambda } \nonumber \\
&& \times \sum_{j=1}^{\infty }\frac{%
\lambda _{n,j}^{(\lambda )}T_{qn}(\lambda _{n,j}^{(\lambda )})}{\sqrt{%
\lambda _{n,j}^{(\lambda )2}+k^{2}a^{2}}}J_{qn}^{2}(\lambda _{n,j}^{(\lambda
)}r/a),  \label{T21}
\end{eqnarray}%
where $i=0,1,2,3$, and we have introduced the notations%
\begin{eqnarray}
f^{(j)}[\Phi (\phi ),f(x)] &=&(-1)^{i}\left( 2k^{2}/\gamma ^{2}+1\right)
\left[ \Phi ^{2}(\phi )f^{\prime 2}(x)+\Phi ^{\prime 2}(\phi )f^{2}(x)/y^{2}%
\right] +\Phi ^{2}(\phi )f^{2}(x),  \notag \\
f^{(l)}[\Phi (\phi ),f(x)] &=&(-1)^{l}\Phi ^{2}(\phi )f^{\prime 2}(x)-\left[
\Phi ^{2}(\phi )+(-1)^{l}\Phi ^{\prime 2}(\phi )/x^{2}\right] f^{2}(x),
\label{fi}
\end{eqnarray}%
with $j=0,3$ and $l=1,2$. As in the case of the field squared, in formulae (%
\ref{Tik}) and (\ref{T21}) we introduce a cutoff function and apply formula (%
\ref{sumform1AP}) for the summation over $n$. This enables us to present the
vacuum energy-momentum tensor in the form of the sum
\begin{equation}
\langle 0|T_{i}^{k}|0\rangle =\langle 0_{w}|T_{i}^{k}|0_{w}\rangle +\langle
T_{i}^{k}\rangle _{\mathrm{cyl}},  \label{Tik1}
\end{equation}%
where $\langle 0_{w}|T_{i}^{k}|0_{w}\rangle $ is the part corresponding to
the geometry of a wedge without a cylindrical boundary and $\langle
T_{i}^{k}\rangle _{\mathrm{cyl}}$ is induced by the cylindrical shell. The
latter may be written in the form (no summation over $i$)
\begin{eqnarray}
\langle T_{i}^{i}\rangle _{\mathrm{cyl}} &=&\frac{q}{2\pi ^{2}}%
\sideset{}{'}{\sum}_{n=0}^{\infty }\sum_{\lambda =0,1}\int_{0}^{\infty
}dxx^{3}\frac{K_{qn}^{(\lambda )}(xa)}{I_{qn}^{(\lambda )}(xa)}F^{(i)}[\Phi
_{qn}^{(\lambda )}(\phi ),I_{qn}(xr)],  \label{Tikb} \\
\langle T_{2}^{1}\rangle _{\mathrm{cyl}} &=&\frac{q^{2}}{4\pi ^{2}}\frac{%
\partial }{\partial r}\sideset{}{'}{\sum}_{n=0}^{\infty }n\sin (2qn\phi
)\sum_{\lambda =0,1}(-1)^{\lambda }\int_{0}^{\infty }dxx\frac{%
K_{qn}^{(\lambda )}(xa)}{I_{qn}^{(\lambda )}(xa)}I_{qn}^{2}(xr),
\label{T21b}
\end{eqnarray}%
with the notations
\begin{eqnarray}
F^{(i)}[\Phi (\phi ),f(y)] &=&\Phi ^{2}(\phi )f^{2}(y),\;i=0,3,  \notag \\
F^{(i)}[\Phi (\phi ),f(y)] &=&-(-1)^{i}\Phi ^{2}(\phi )f^{\prime 2}(y)-\left[
\Phi ^{2}(\phi )-(-1)^{i}\Phi ^{\prime 2}(\phi )/y^{2}\right]
f^{2}(y),\;i=1,2.  \label{Fnui}
\end{eqnarray}%
The diagonal components are symmetric with respect to the half-plane $\phi
=\phi _{0}/2$, whereas the off-diagonal component is an odd function under
the replacement $\phi \rightarrow \phi _{0}-\phi $. As it can be easily
checked, the tensor $\langle T_{i}^{k}\rangle _{\mathrm{cyl}}$ is traceless
and satisfies the covariant continuity equation. The off-diagonal component $%
\langle T_{2}^{1}\rangle _{\mathrm{cyl}}$ vanishes at the wedge sides and
for these points the VEV of the energy-momentum tensor is diagonal. The
vacuum energy density induced by the cylindrical shell in the interior
region is always negative.

The renormalized VEV of the energy density for the geometry without the
cylindrical shell is obtained by using the corresponding formulae for the
field squared. Other components are found from the tracelessness condition
and the continuity equation and one has \cite{Most97,jphy,Deutsch}
\begin{equation}
\langle T_{i}^{k}\rangle _{\mathrm{ren}}^{(w)}=-\frac{(q^{2}-1)(q^{2}+11)}{%
720\pi ^{2}r^{4}}\mathrm{diag}(1,1,-3,1).  \label{Tikw}
\end{equation}%
Formula (\ref{Tikw}) coincides with the corresponding result for the
geometry of the cosmic string with the angle deficit $2\pi -\phi _{0}$ and
in the corresponding formula $q=2\pi /\phi _{0}$.

The normal force acting on the wedge sides is determined by the component $%
\langle T_{2}^{2}\rangle _{\mathrm{ren}}$ of the vacuum energy-momentum
tensor evaluated for $\phi =0$ and $\phi =\phi _{0}$. On the base of formula
(\ref{Tik1}) for the corresponding effective pressure one has%
\begin{equation}
p_{2}=-\langle T_{2}^{2}\rangle _{\mathrm{ren}}|_{\phi =0,\phi
_{0}}=p_{2w}+p_{2\mathrm{cyl}},  \label{p2}
\end{equation}%
where $p_{2w}=-\langle T_{2}^{2}\rangle _{\mathrm{ren}}^{(w)}$ is the normal
force acting per unit surface of the wedge for the case without a
cylindrical boundary and the additional term%
\begin{equation}
p_{2\mathrm{cyl}}=-\langle T_{2}^{2}\rangle _{\mathrm{cyl}}|_{\phi =0,\phi
_{0}}=-\frac{q}{\pi ^{2}}\sideset{}{'}{\sum}_{n=0}^{\infty }\sum_{\lambda
=0,1}\int_{0}^{\infty }dxx^{3}\frac{K_{qn}^{(\lambda )}(xa)}{%
I_{qn}^{(\lambda )}(xa)}F_{qn}^{(\lambda )}[I_{qn}(xr)],  \label{p2cyl}
\end{equation}%
with the notations%
\begin{equation}
F_{\nu }^{(0)}[f(y)]=\nu ^{2}f^{2}(y)/y^{2},\;F_{\nu
}^{(1)}[f(y)]=-f^{^{\prime }2}(y)-f^{2}(y),  \label{Fnulam}
\end{equation}%
is induced by the cylindrical shell. Note that the normal force on the wedge
sides is the sum of the corresponding forces for Dirichlet and Neumann
scalars corresponding to the terms with $\lambda =0$ and $\lambda =1$
respectively. The finiteness of the normal stress on the wedge sides is a
consequence of the fact that for a single perfectly conducting plane
boundary this stress vanishes. Note that this result can be directly
obtained from the symmetry of the corresponding problem with combination of
the continuity equation for the energy-momentum tensor. It also survives for
more realistic models of the plane boundary (see, for instance, \cite{Helf99}%
) though the corresponding energy density and parallel stresses no longer
vanish. So we expect that the obtained formula for the normal force acting
on the wedge sides will correctly approximate the corresponding results of
more realistic models in the perfectly conducting limit. The corresponding
vacuum forces are attractive for $q>1$ and repulsive for $q<1$. In
particular, the equilibrium position corresponding to the geometry of a
single plate ($q=1$) is unstable. As regards to the part induced by the
cylindrical shell, from (\ref{p2cyl}) it follows that $p_{2\mathrm{cyl}}<0$
and, hence, the corresponding forces are always attractive. In figure \ref%
{fig2} we have plotted the vacuum pressure on the wedge sides induced by the
cylindrical boundary versus $r/a$ for Dirichlet scalar (left panel) and for
the electromagnetic field (right panel). The full (dashed) curves correspond
to the wedge with $\phi _{0}=\pi /2$ ($\phi _{0}=3\pi /2$).
\begin{figure}[tbph]
\begin{center}
\begin{tabular}{cc}
\epsfig{figure=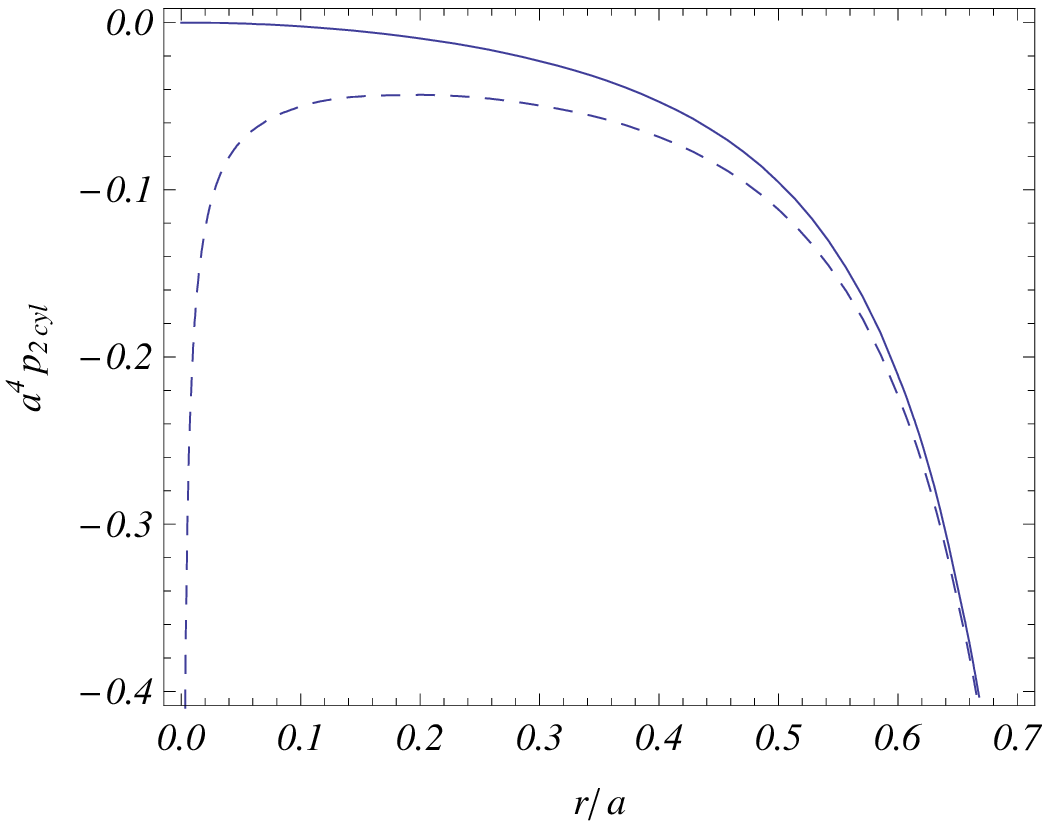,width=7cm,height=5.5cm} & \quad %
\epsfig{figure=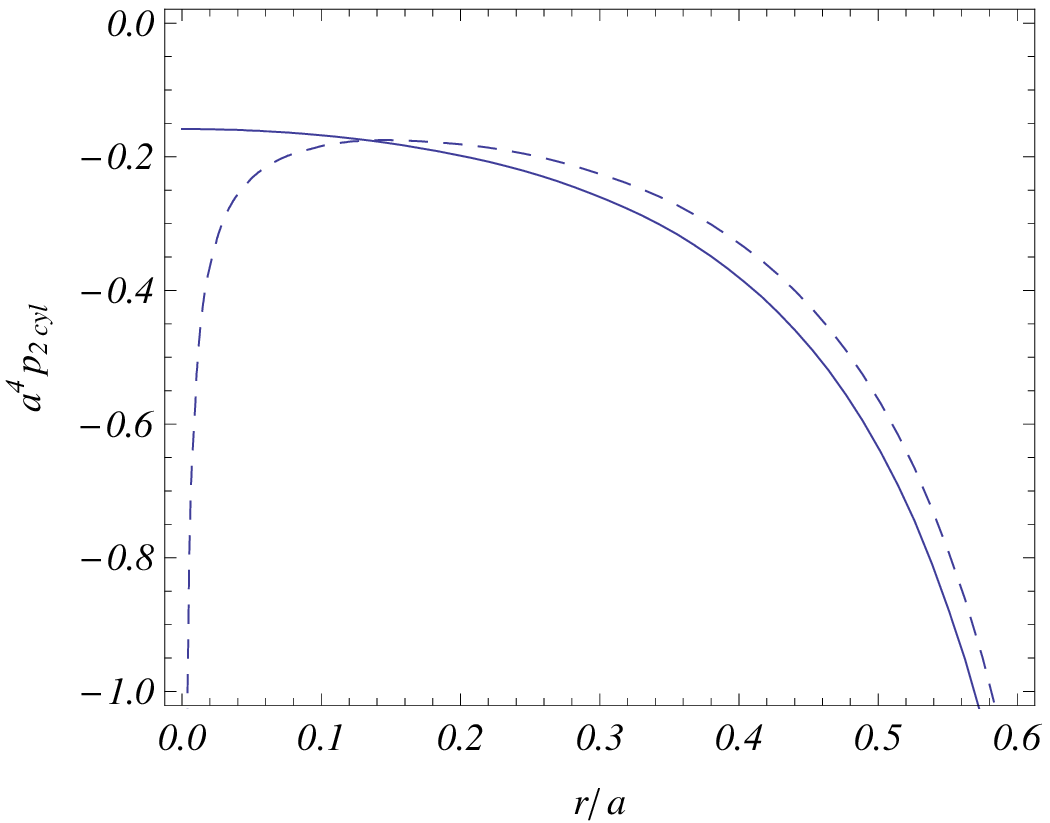,width=7cm,height=5.5cm}%
\end{tabular}%
\end{center}
\caption{The effective azimuthal pressure induced by the cylindrical shell
on the wedge sides, $a^{4}p_{2\mathrm{cyl}}$, as a function of $r/a$ for
Dirichlet scalar (left panel) and for the electromagnetic field (right
panel). The full (dashed) curves correspond to $q=2$ ($q=2/3$).}
\label{fig2}
\end{figure}

Now, let us discuss the behavior of the boundary-induced part in the VEV of
the energy-momentum tensor in the asymptotic regions of the parameters. Near
the cylindrical shell the main contribution comes from large values of $n$
and for the points $a-r\ll a|\sin \phi |,a|\sin (\phi _{0}-\phi )|$ the
leading terms are the same as those for a cylindrical shell when the wedge
is absent. For points near the edges $(r=a,\phi =0,\phi _{0})$ the leading
terms in the corresponding asymptotic expansions are the same as for the
geometry of a wedge with the opening angle $\phi _{0}=\pi /2$. Near the
edge, $r\rightarrow 0$, for the components (no summation over $i$) $\langle
T_{i}^{i}\rangle _{\mathrm{cyl}}$, $i=0,3$, the main contribution comes from
the mode $n=0$ and we find $\langle T_{i}^{i}\rangle _{\mathrm{cyl}}\approx
-0.0590q/a^{4}$, $i=0,3$. For the components (no summation over $i$) $%
\langle T_{i}^{i}\rangle _{\mathrm{cyl}}$, $i=1,2$, when $q>1$ the main
contribution again comes form $n=0$ term and one has $\langle
T_{i}^{i}\rangle _{\mathrm{cyl}}\approx -\langle T_{0}^{0}\rangle _{\mathrm{%
cyl}}$, $i=1,2$. For $q<1$ the main contribution into the components $%
\langle T_{i}^{i}\rangle _{\mathrm{cyl}}$, $i=1,2$, comes from the term $n=1$
and we have (no summation over $i$) $\langle T_{i}^{i}\rangle _{\mathrm{cyl}%
}\varpropto r^{2(q-1)}$, $i=1,2$. In this case the radial and azimuthal
stresses induced by the cylindrical shell diverge on the edge $r=0$. For the
off-diagonal component the main contribution comes from the $n=1$ mode and $%
\langle T_{2}^{1}\rangle _{\mathrm{cyl}}$ behaves like $r^{2q-1}$.

Now we turn to the VEVs of the energy-momentum tensor in the exterior
region. Subtracting from these VEVs the corresponding expression for the
wedge without the cylindrical boundary, analogously to the case of the field
square, it can be seen that the VEVs are presented in the form (\ref{Tik1}),
with the parts induced by the cylindrical shell given by the formulae (no
summation over $i$)
\begin{eqnarray}
\langle T_{i}^{i}\rangle _{\mathrm{cyl}} &=&\frac{q}{2\pi ^{2}}%
\sideset{}{'}{\sum}_{n=0}^{\infty }\sum_{\lambda =0,1}\int_{0}^{\infty
}dxx^{3}\frac{I_{qn}^{(\lambda )}(xa)}{K_{qn}^{(\lambda )}(xa)}F^{(i)}[\Phi
_{qn}^{(\lambda )}(\phi ),K_{qn}(xr)],  \label{Tikbext} \\
\langle T_{2}^{1}\rangle _{\mathrm{cyl}} &=&\frac{q^{2}}{4\pi ^{2}}\frac{%
\partial }{\partial r}\sideset{}{'}{\sum}_{m=0}^{\infty }n\sin (2qn\phi
)\sum_{\lambda =0,1}(-1)^{\lambda }\int_{0}^{\infty }dxx\frac{%
I_{qn}^{(\lambda )}(xa)}{K_{qn}^{(\lambda )}(xa)}K_{qn}^{2}(xr).
\label{T21bext}
\end{eqnarray}%
Here the functions $F^{(i)}\left[ \Phi (\phi ),f(y)\right] $ are
defined by expressions (\ref{Fnui}). It can be seen that the
vacuum energy density induced by the cylindrical shell in the
exterior region is positive.

In the way similar to that for the interior region, the force acting on the
wedge sides is presented in the form of the sum (\ref{p2}), where for the
part due to the presence of the cylindrical shell we have%
\begin{equation}
p_{2\mathrm{cyl}}=-\langle T_{2}^{2}\rangle _{\mathrm{cyl}}|_{\phi =0,\phi
_{0}}=-\frac{q}{\pi ^{2}}\sideset{}{'}{\sum}_{n=0}^{\infty }\sum_{\lambda
=0,1}\int_{0}^{\infty }dxx^{3}\frac{I_{qn}^{(\lambda )}(xa)}{%
K_{qn}^{(\lambda )}(xa)}F_{qn}^{(\lambda )}[K_{qn}(xr)].  \label{p2cylext}
\end{equation}%
In this formula, the function $F_{\nu }^{(\lambda )}\left[ f(y)\right] $ is
defined by relations (\ref{Fnulam}) and the corresponding forces are always
attractive.

The leading divergence in the boundary induced part (\ref{Tikbext}) on the
cylindrical surface is given by the same formulae as for the interior
region. For large distances from the shell and for $q>1$ the main
contribution into the VEVs of the diagonal components comes from the $n=0$, $%
\lambda =1$ term and one has (no summation over $i$)%
\begin{equation}
\langle T_{i}^{i}\rangle _{\mathrm{cyl}}\approx -\frac{qc_{i}\left(
a/r\right) ^{2}}{15\pi ^{2}r^{4}},\;c_{0}=c_{3}=2,\;c_{1}=1,\;c_{2}=-5.
\label{Tikfar}
\end{equation}%
In the case $q<1$ the main contribution into the VEVs of the diagonal
components at large distances from the cylindrical shell comes from the $n=1$
mode. The leading terms in the corresponding asymptotic expansions are given
by the formulae%
\begin{equation}
\langle T_{i}^{i}\rangle _{\mathrm{cyl}}\approx -q^{2}(q+1)c_{i}(q)\frac{%
\cos (2q\phi )}{\pi ^{2}r^{4}}\left( \frac{a}{r}\right) ^{2q},
\label{Tiifar1}
\end{equation}%
with the notations%
\begin{equation}
c_{0}(q)=c_{3}(q)=\frac{1}{2q+3},\;c_{1}(q)=\frac{2q^{2}+q+1}{(2q+1)(2q+3)}%
,\;c_{2}(q)=-\frac{q+1}{2q+1}.  \label{ci(q)}
\end{equation}%
In the case $q=1$ the asymptotic terms are determined by the sum of the
contributions coming from the modes $n=0$ and $n=1$. The latter are given by
formulae (\ref{Tikfar}), (\ref{Tiifar1}). For the off-diagonal component,
for all values $q$ the main contribution at large distances comes from the $%
n=1$ mode and $\langle T_{2}^{1}\rangle _{\mathrm{cyl}}\varpropto
(a/r)^{2q}r^{-3}$.

\section{Conclusion}

\label{sec:Conc}

We have investigated the polarization of the scalar and electromagnetic
vacua by a wedge with coaxial cylindrical boundary, assuming Dirichlet
boundary conditions in the case of a scalar field and perfectly conducting
boundary conditions for the electromagnetic field. The application of the
Abel-Plana-type summation formula for the series over the zeros of the
Bessel function and its derivative allowed us to extract from the VEVs the
parts due to the wedge without a cylindrical boundary and to present the
additional parts induced by this boundary in terms of exponentially
convergent integrals. The vacuum densities for the geometry of a wedge
without a cylindrical boundary are considered in section \ref%
{sec:Wedgecylabs}. We have derived formulae for the renormalized VEVs of the
field squared and the energy-momentum tensor, formulae (\ref{phi2w}), (\ref%
{Tiiw21}). For a conformally coupled scalar the energy-momentum tensor is
diagonal and does not depend on the angular variable $\phi $. The
corresponding vacuum forces acting on the wedge sides are attractive for $%
\phi _{0}<\pi $ and are repulsive for $\phi _{0}>\pi $.

For a scalar field the parts in the Wightman function induced by the
cylindrical boundary are given by formulae (\ref{Wfa0}) and (\ref{Wfa0ext})
for the interior and exterior regions respectively. The corresponding VEVs
for the field squared and the energy-momentum tensor are investigated in
section \ref{sec:Wedgecyl}. The field squared is given by formula (\ref%
{phi2a1}) and vanishes on the wedge sides $\phi =\phi _{m}$ for all points
away from the cylindrical surface. The energy-momentum tensor induced by the
cylindrical surface is non-diagonal and the corresponding components are
determined by formulae (\ref{Tiia}), (\ref{Tiia21}). The off-diagonal
component vanishes on the wedge sides. The additional vacuum forces acting
on the wedge sides due to the presence of the cylindrical surface are
determined by the $_{2}^{2}$-component of the corresponding stress and are
attractive for all values $\phi _{0}$. On the wedge sides the corresponding
vacuum stresses in the directions parallel to the wedge sides are isotropic
and the energy density is negative for both minimally and conformally
coupled scalars. The formulae in the exterior region differ from the
corresponding formulae for the interior region by the interchange $%
I_{qn}(z)\leftrightarrows K_{qn}(z)$. For large distances from the
cylindrical surface, $r\gg a$, the VEVs behave as $(a/r)^{D-1+2q}$ for the
field squared and as $(a/r)^{D+1+2q}$ for the diagonal components of the
energy-momentum tensor.

In the second part of the paper we have evaluated the VEVs of the field
squared and the energy-momentum tensor for the electromagnetic field. For
the wedge without the cylindrical shell the VEVs of the field squared are
presented in the form (\ref{F2wren}). The first term on the right of this
formula corresponds to the VEVs for the geometry of a cosmic string with the
angle deficit $2\pi -\phi _{0}$. The parts induced by the cylindrical shell
are presented in the form (\ref{F2b0}) for the interior region and in the
form (\ref{F2bext}) for the exterior region. We have discussed these general
formulae in various asymptotic regions of the parameters. In section \ref%
{sec:EMTint} we consider the VEV of the energy-momentum tensor.
For the geometry of a wedge without the cylindrical boundary the
vacuum energy-momentum tensor\ does not depend on the angle $\phi
$ and is the same as in the geometry of the cosmic string. The
corresponding vacuum forces acting on the wedge sides are
attractive for $\phi _{0}<\pi $ and repulsive for $\phi _{0}>\pi
$. In particular, the equilibrium position corresponding to the
geometry of a single plate is unstable. For the region inside the
shell the part in the VEV of the energy-momentum tensor induced by
the presence of the cylindrical shell is non-diagonal and the
corresponding components are given by formulae (\ref{Tikb}),
(\ref{T21b}) for the interior region and by (\ref{Tikbext}),
(\ref{T21bext}) for the exterior region. The vacuum energy density
induced by the cylindrical shell is negative in the
interior region and is positive in the exterior region. For a wedge with $%
\phi _{0}<\pi $ the part in the vacuum energy-momentum tensor induced by the
shell is finite on the edge $r=0$. For $\phi _{0}>\pi $ the shell-induced
parts in the energy density and the axial stress remain finite, whereas the
radial and azimuthal stresses diverge as $r^{2(\pi /\phi _{0}-1)}$. The
corresponding off-diagonal component behaves like $r^{2\pi /\phi _{0}-1}$
for all values $\phi _{0}$. For points near the edges $(r=a,\phi =0,\phi
_{0})$, the leading terms in the corresponding asymptotic expansions are the
same as for the geometry of a wedge with the opening angle $\phi _{0}=\pi /2$%
. The presence of the shell leads to additional forces acting on the wedge
sides. The corresponding effective azimuthal pressures are given by formulae
(\ref{p2cyl}), (\ref{p2cylext}) and these forces are always attractive.

\section*{Acknowledgments}

The work was supported by the Armenian Ministry of Education and Science
Grant No. 119.

\end{document}